\documentclass{lamuphys}
\usepackage{graphics}
\begin{document}
\title{Black Holes: A General Introduction}
\author{Jean-Pierre Luminet} 
\institute{Observatoire de Paris-Meudon,
D\'epartement d'Astrophysique Relativiste et de  Cosmologie,
CNRS UPR-176, F-92195 Meudon Cedex, France}
\maketitle
\label{chap:luminet}

\begin{abstract}
Our understanding of space and time is probed to its depths by black holes.
These objects, which appear as a natural consequence of general relativity,
provide a powerful analytical tool able to examine macroscopic and
microscopic properties of the universe. This introductory article presents
in a pictorial way the basic concepts of black hole's theory, as well as a
description of the astronomical sites where black holes are suspected to
lie, namely binary X--ray sources and galactic nuclei.
\end{abstract}

\section {The Black Hole Mystery}

\index{butterflies}
Let me begin with an old Persian story. Once upon a time, the butterflies
organized a summer school devoted to the great
mystery of the flame. Many discussed about models but nobody could
convincingly explain the puzzle. Then a bold butterfly enlisted as a
volunteer to get a real experience with the flame. He flew off to the
closest castle, passed in front of a window and saw the light of a candle. He
went back, very excited, and told what he had seen. But the wise 
butterfly who was the chair of the conference said that they had no
more information than before.
Next, a second butterfly flew off to the castle, crossed the window and
touched the flame with his wings. He hardly came back and told his story; the wise
chairbutterfly said ``your explanation is no more satisfactory". Then a third
butterfly went to the castle, hit the candle and burned himself into the
flame. The wise butterfly, who had observed the action, said to the others:
``Well, our friend has learned everything about the flame. But only him can
know, and that's all".

As you can guess, this story can easily be transposed from butterflies to
scientists confronted with the mystery of black holes. Some
astronomers, equipped with powerful instruments such as orbiting telescopes,
make very distant and indirect observations on black holes; like the first
butterfly, they acknowledge the real existence of black holes but they gain
very little information on their real nature. Next, theoretical physicists try
to penetrate more deeply into the black hole mystery by using tools such as
general relativity, quantum mechanics and higher mathematics; like the second
butterfly, they get a little bit more information, but not so much. The
equivalent of the third butterfly would be a spationaut plunging directly into
a black hole, but eventually he will not be able to go back and tell his story.
Nevertheless, by using numerical calculations such as those performed at the
Observatoire de Meudon I will show you later, outsiders can get some idea of what
happens inside a black hole. 

\section {Physics of Black Holes}

\subsection {Light imprisoned}
 
\index{black hole!definition}
Let us begin to play like the second butterfly, and explore the black hole 
from  the point of view of theoretical physics.
An elementary definition of a black hole is a region of space-time in which
the gravitational potential, $GM/R$, exceeds the square of the speed of light,
$c^2$. Such a statement has the merit to be independent of the details of
gravitational theories. It can be used in the framework of Newtonian
theory. It also provides a more popular definition of a black
hole, according to which any astronomical body whose
escape velocity exceeds the speed of light must be a black hole.
Indeed, such a reasoning was done two centuries ago
by John Michell and Pierre-Simon de Laplace. In the {\it Philosophical
Transactions of the Royal Society} (1784), John Michell pointed out that ``if
the semi diameter of a sphere of the same density with the sun were to exceed
that of the sun in the proportion of 500 to 1, (...) all light emitted from
such a body would be made to return towards it", and independently, in 1796,
Laplace wrote in his {\it Exposition du Syst\`eme du Monde}: ``Un astre
lumineux de m\^eme densit\'e que la terre et dont le diam\`etre serait deux cents
cinquante fois plus grand que celui du soleil, ne laisserait, en vertu de son
attraction, parvenir aucun de ses rayons jusqu'\`a nous ; il est donc possible
que les plus grands corps lumineux de l'univers soient, par cela m\^eme,
invisibles".  Since the density imagined at this time was that of ordinary
matter, the size and the mass of the associated ``invisible body" were huge -
around $10^7$ solar masses, corresponding to what is called today a
``supermassive" black hole. Nevertheless, from the numerical figures first
proposed by Michell and Laplace, one can recognize the well-known basic
formula giving the critical radius of a body of mass $M$: 
\begin{equation}
R_S = {2 GM\over c^2} \approx 3 {M\over M_{\odot}} \textrm{km},     
\end{equation}
where $M_{\odot}$ is the solar mass. Any spherical body of mass $M$ confined 
within the critical radius $R_{S}$ must be a black hole.
 
These original speculations were quickly forgotten, mainly due to the
development of the wave theory of light, within the framework of which no
calculation of the action of gravitation on light propagation was performed.
The advent of general relativity, a fully relativistic
theory of gravity in which light is submitted to gravity, gave rise to new
speculations and much deeper insight into black holes.

To pictorially describe black holes in space-time, I shall use light cones.
Let me recall what a light cone is.
In figure 1, a luminous flash is emitted at a given point of
space. The wavefront is a sphere expanding at a velocity of 
$c = 300\;000 \, km/s$,
shown in a) at three successive instants. The light cone
representation in b) tells the complete story of the wavefront in a single
space-time diagram. As one space dimension is removed, the spheres become
circles. The expanding circles of light generate a cone originating at the
emission point. If, in this diagram, we choose the unit of length as 300 000 km
and the unit of time as 1 second, all the light rays
travel at $45^{\circ}$.

\begin{figure}[htb]
  \begin{center}
    \leavevmode
    \includegraphics{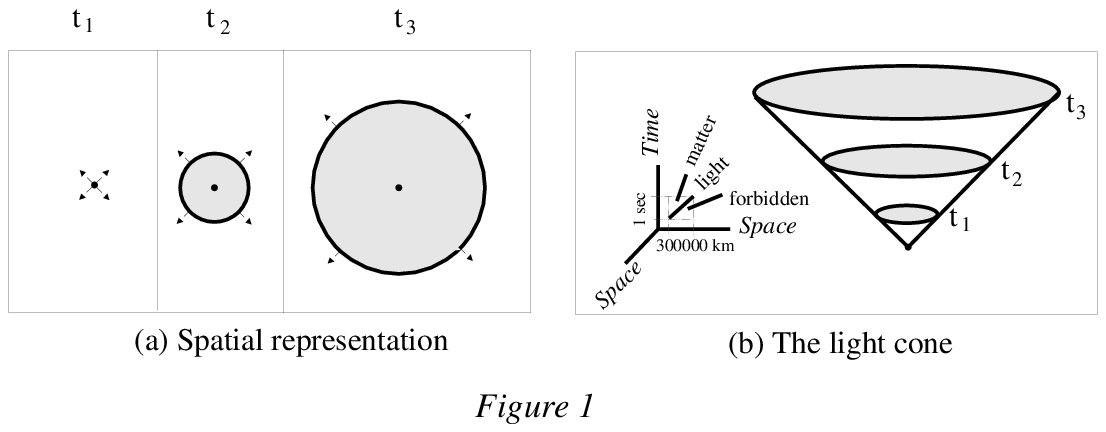}
    \caption{\textbf{The light cone.}}
  \end{center}
\end{figure}
The light cone allows us to depict the causal structure of any space-time. Take
for instance the Minkowski flat space-time used in Special
Relativity (figure 2). At any event $E$ of space-time, light rays generate two
cones (shaded zone). The rays emitted from $E$ span the future light cone, those
received in $E$ span the past light cone.  Physical particles cannot travel
faster than light: their trajectories remain confined within the light
cones.
\begin{figure}[htb]
\renewcommand \thefigure {2\&3}
  \begin{center}
    \leavevmode
    \includegraphics{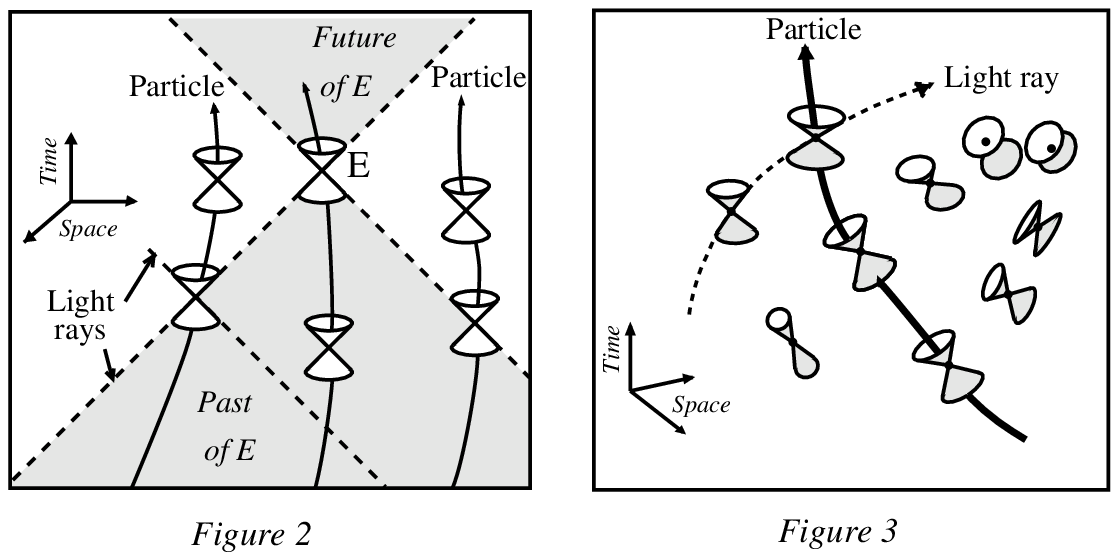} 
    \caption{The space-time continuum of Special Relativity and
             the soft space-time of General Relativity.}
\end{center}
\end{figure}
\setcounter{figure}{3}
No light ray or particle which passes through $E$ is able to penetrate the
clear zone. The invariance of the speed of light in vacuum is
reflected by the fact that all the  cones have the same slope. This is because
the space-time continuum of Special Relativity, free from gravitating matter,
is flat and rigid. As soon as gravity is present,
space-time is curved and Special Relativity leaves room to General Relativity.
Since the 
\index{equivalence principle}Equivalence Principle
states the influence of gravity on all types of energy,
the light cones follow the curvature of space-time (figure 3). They bend
and deform themselves according to the curvature. Special Relativity
remains locally valid however: the worldlines of material
particles remain confined within the light cones, even when the latter are
strongly tilted and distorted by gravity. 

\subsection {Spherical collapse}
\index{collapse!spherical}

Let us now
examine the causal structure of space-time around a gravitationally collapsing
star - a process which is believed to lead to black hole formation. 
\begin{figure}[tb]
  \begin{center}
    \leavevmode
    \includegraphics{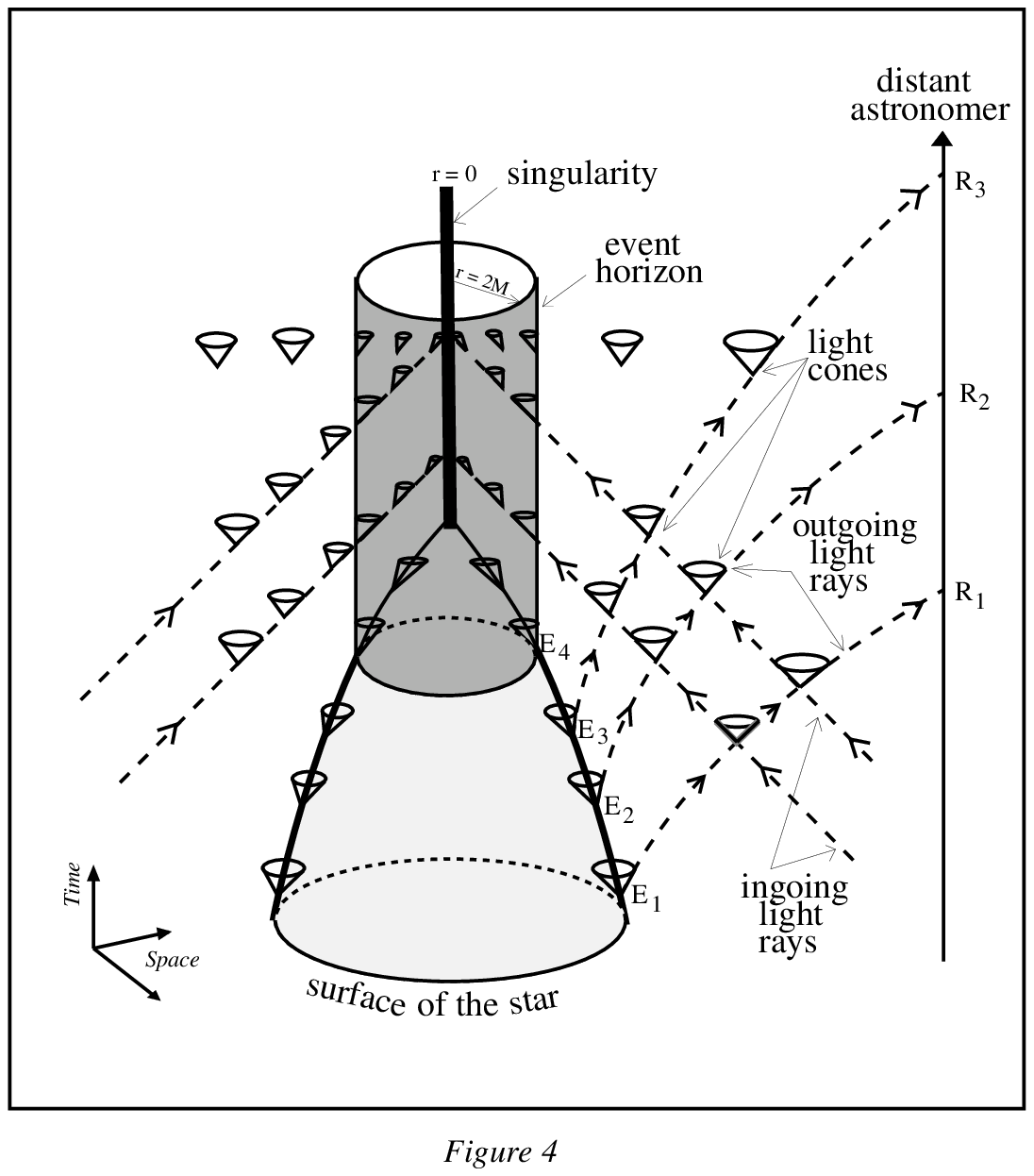}
    \caption{A space-time diagram showing the formation of a black hole by
      gravitational collapse.}
  \end{center}
\end{figure}
Figure 4 shows the complete history of the collapse of a spherical 
star, from
its initial contraction until the formation of a black hole and a singularity.
Two
space dimensions are measured horizontally, and time is on the vertical axis,
measured upwards. The centre of the star is at $r=0$.  The curvature of
space-time is visualized by means of the light cones generated by the
trajectories of light rays. Far away from the central gravitational field, the
curvature is so weak that the light cones remain straight. Near the
gravitational field, the cones are distorted and tilted inwards by the curvature. On
the critical surface of radius $r=2M$, the cones are tipped over at 
$45^{\circ}$
and one of their generators becomes vertical, so that the allowed directions of
propagation of particles and electromagnetic waves are oriented towards 
the interior of this surface. This is the {\it event horizon}, the boundary of the
black hole (grey region). Beyond this, the stellar matter continues to
collapse into a singularity of zero volume and infinite density at $r=0$. Once
a black hole has formed, and after all the stellar matter has disappeared into
the singularity, the geometry of space-time itself continues to collapse
towards the singularity, as shown by the light cones. 

The emission of the light
rays at $E_1, E_2, E_3$ and $E_4$ and their reception by a distant astronomer
at $R_1, R_2, R_3, ...$ well illustrate the difference between the {\it proper
time}, as measured by a clock placed on the surface of the star, and the {\it
apparent time}, measured by an independent and distant clock. The (proper) time
interval between the four emission events are equal. However, the
corresponding reception intervals become longer and longer. At the limit,
light ray emitted from $E_4$, just when the event horizon is forming, takes an
infinite time to reach the distant astronomer. This phenomenon of ``frozen
time" is just an illustration of the extreme elasticity of time predicted by
Einstein's relativity, according to which time runs differently for two
observers with a relative acceleration - or, from the Equivalence Principle,
in different gravitational potentials. A striking consequence is that any
outer astronomer will {\it never} be able to see the formation of a black
hole. The figure 5 shows a picturesque illustration of frozen time. A
spaceship has the mission of exploring the interior of a black hole --
preferably a big one, so that it is not destroyed too quickly by the tidal
forces. On board the ship, the commander sends a solemn salute to mankind,
just at the moment when the ship crosses the horizon. His gesture is
transmitted to distant spectators via television. The film on the left shows
the scene on board the spaceship in proper time, that is, as measured by the
ship's clock as the ship falls into the black hole. The astronaut's salute
is decomposed into instants at proper time intervals of 0.2 second. Crossing of
the event horizon (black holes have not a {\it hard} surface)  is not
accompanied by any particular event. The film on the right shows the scene
received by distant spectators via television. It is also decomposed
into intervals of apparent time of 0.2 second. At the beginning of his
gesture, the salute is slightly slower than the real salute, but initially
the delay is too small to be noticed, so the films are practically
identical. It is only very close to the horizon that apparent time starts
suddenly to freeze;  the film on the right then shows the astronaut eternally
frozen in the middle of his salute, imperceptibly reaching the limiting
position where he crossed the horizon. Besides this effect, the shift in the
frequencies in the gravitational field  (the so-called Einstein's effect)
causes the images to weaken, and they soon become invisible.
\begin{figure}[tb]
  \begin{center}
    \leavevmode
    \includegraphics{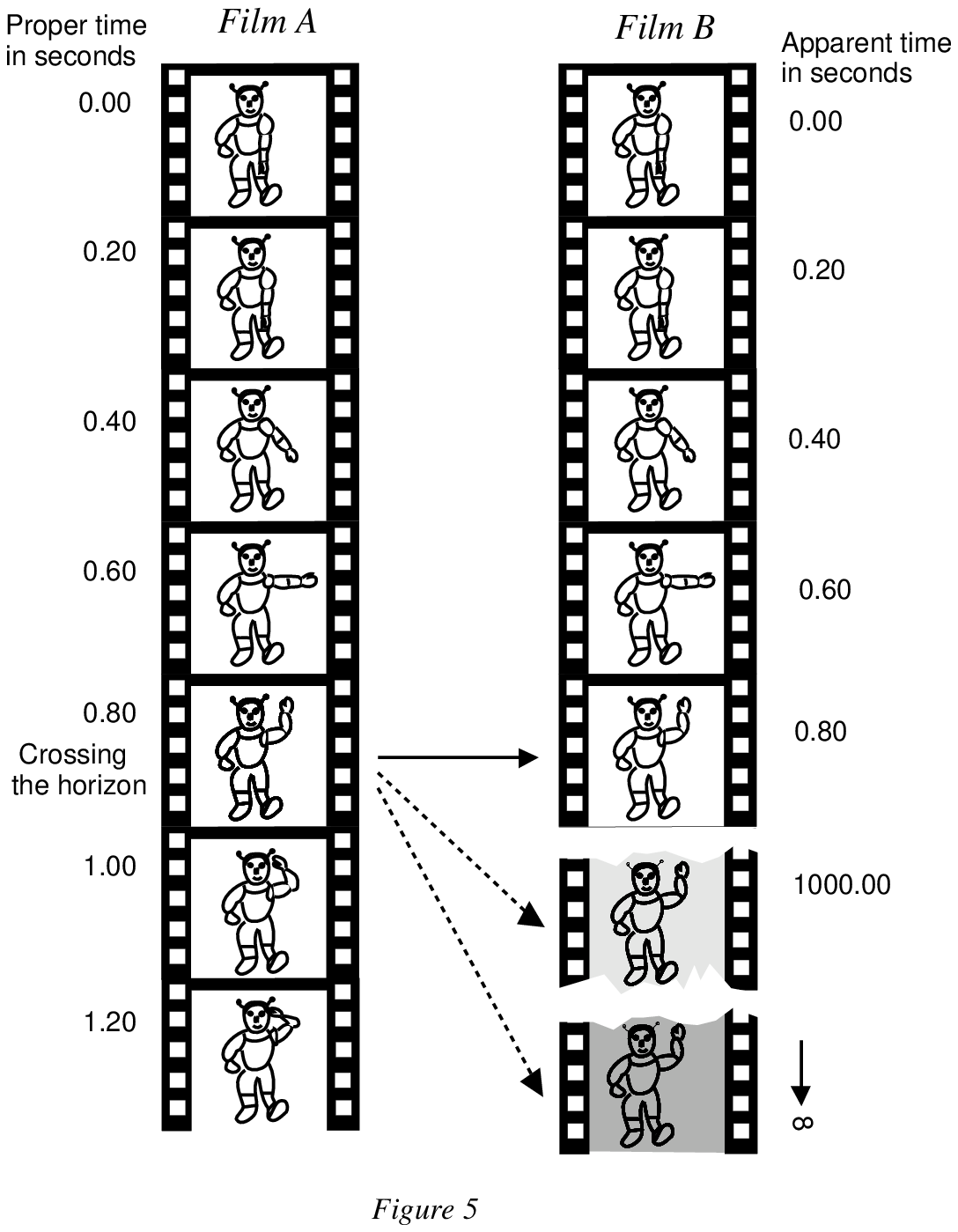}
    \caption{The astronauts salute.}
  \end{center}
\end{figure}

All these effects follow rather straightforwardly from equations.
In General Relativity, the vacuum space-time around a spherically symmetric body
is described by the \index{Schwarzschild solution}Schwarzschild metric: 
\begin{equation}
ds^2 = - \left(1- \frac{2M}{r}\right)dt^2 
+ \left(1-\frac{2M}{r}\right)^{-1}dr^2 + r^2 d\Omega^2,
\end{equation}
where $d\Omega^2 = d\theta^2 + sin^2\theta \;d\phi^2$ is the metric of a unit
2-sphere, and we have set the gravity's constant $G$ and the speed of light $c$
equal to unity. The solution describes the external gravitational field
generated by any static spherical mass, whatever its radius (Birkhoff's theorem,
1923). 

When the radius is greater than $2M$, there exists ``interior solutions"
depending on the equation of state of the stellar matter, which are
non-singular at $r=0$ and that match the exterior solution.  However,
as soon as the body is collapsed under its critical radius $2M$, the
Schwarzschild metric is the unique solution for the gravitational
field generated by a spherical black hole. The event horizon, a sphere of radius
$r=2M$, is a coordinate singularity which can be removed by a suitable
coordinate transformation (see below). There is a true gravitational
singularity at $r=0$ (in the sense that some curvature components diverge) that
cannot be removed by any coordinate transformation. Indeed the singularity does
not belong to the space-time manifold itself. 
\begin{figure}[tb]
  \begin{center}
    \leavevmode
    \includegraphics{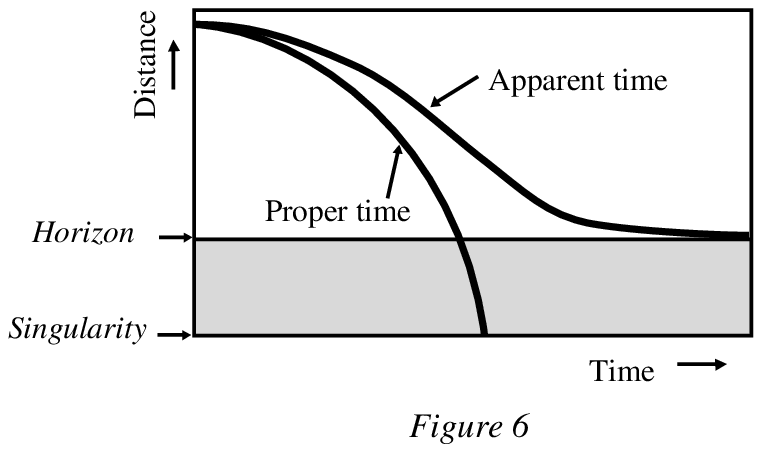}
    \caption{The two times of a black hole.}
  \end{center}
\end{figure}
Inside the event horizon, the radial coordinate $r$ becomes timelike. Hence
every particle that crosses the event horizon is unavoidably catched by the
central singularity. For radial free-fall along a trajectory with  $r
\rightarrow 0 $, the proper time (as measured by a comoving clock) is given by 
\begin{equation}
\tau =  \tau_0 - \frac{4M}{3}\left(\frac{r}{2M}\right)^{3/2}
\end{equation}
and is well-behaved at the event horizon.
 The apparent time (as measured by a distant observer) is given by  
\begin{equation}
t = \tau - 4M\left(\frac{r}{2M}\right)^{1/2}
+ 2M \ln\frac{\sqrt{r/2M}+1}{\sqrt{r/2M}-1},
\end{equation}
and diverges to infinity as $r \to 2M$, see figure 6.

The Schwarzschild coordinates, which cover only 
$2M \le r < \infty , - \infty < t < + \infty $, are not well adapted to the 
analysis of the causal structure of
space-time near the horizon,  because the light cones, given by $dr = \pm
(1-\frac{2M}{r})dt$, are not defined on the event horizon. We better use the
so-called Eddington-Finkelstein coordinates - indeed discovered by 
Lema\^{\i}tre in
1933 but remained unnoticed. Introducing the ``ingoing" coordinate 
\begin{equation}
v = t+r+2Mln(\frac{r}{2M}-1)    
\end{equation}
 the Schwarzschild metric becomes 
\begin{equation}
 ds^2 = - (1-\frac{2M}{r})dv^2 + 2 dvdr + r^2d\Omega^2.
\end{equation}
Now the light cones are perfectly well behaved. The ingoing light rays are
given by 
\begin{equation}
dv = 0, 
\end{equation}
the outgoing light rays by 
\begin{equation}
dv = \frac{2dr}{1-\frac{2M}{r}}.
\end{equation}
The metric can be analytically continued to all $r>0$ and
is no more singular at $r=2M$.
Indeed, in figure 4 such a coordinate system was already used.

\subsection {Non spherical collapse}
A black hole may well form from an asymmetric gravitational collapse. However
the deformations of the event horizon are quickly dissipated as 
gravitational radiation; the event horizon vibrates according to the so--called
``quasi-normal modes" and the black hole settles down into a final axisymmetric
equilibrium configuration.  

\begin{figure}[tb]
  \begin{center}
    \leavevmode
    \includegraphics{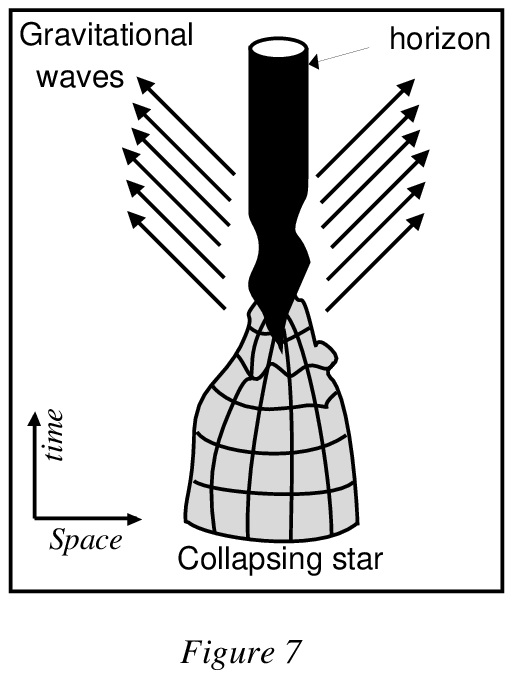}       
    \caption{Gravitational collapse of a star.}
  \end{center}
\end{figure}
The deepest physical property of black holes is that asymptotic
equilibrium solutions depend only on three parameters:
the mass, the electric charge and the angular momentum. All the
details of the infalling matter other than mass, electric charge and 
angular momentum are washed out. The proof followed from efforts
over 15 years by half a dozen of theoreticians, but it was originally suggested as
a conjecture by John Wheeler, who used the picturesque formulation ``a black
hole has no hair".
Markus Heusler's lectures in this volume will develop this so--called
``uniqueness theorem".  

As a consequence, there exists only 4 exact solutions of Einstein's equations
describing black hole solutions with or without charge and angular momentum:

 \begin{itemize}  
\item The Schwarzschild solution (1917) has only mass $M$; it is static,
spherically symmetric.
  \item  The Reissner-Nordstr\"om solution (1918),
static, spherically symmetric, depends on mass $M$ and electric charge $Q$. 
\item The Kerr solution (1963), stationary, axisymmetric,  depends on mass and
angular momentum.
\item The Kerr-Newman solution (1965),  stationary and axisymmetric, depends on
all three parameters $M, J, Q$. 
\end{itemize}
The 3-parameters Kerr-Newman family is the most general solution,
corresponding to the final state of black hole equilibrium. In
Boyer-Lindquist coordinates, the Kerr-Newman metric is given
by

\begin{eqnarray}
        \lefteqn{ds^2  = - (1- \frac{2Mr}{\Sigma})dt^2 
        - 4Mra \frac{sin^2\theta}{\Sigma}dtd\phi}\nonumber\\
 & + (r^2 + a^2 + \frac{2Mr a^2 sin^2\theta}{\Sigma})sin^2\theta d\phi^2 + \frac{
\Sigma}{\Delta}dr^2 + \Sigma d\theta^2  \label{luminet:KN}
\end{eqnarray}
where
$\Delta \equiv r^2 - 2 Mr + a^2 + Q^2$, $ \Sigma \equiv r^2 + a^2 cos^2\theta$, 
$a \equiv J/M$ is the angular momentum per unit mass.
The event horizon is located at distance $r_+ = M + \sqrt{M^2 - Q^2 - a^2}$.
 
From this formula we can see, however, that the black hole parameters cannot
be arbitrary.
Electric charge and angular momentum cannot exceed values corresponding to
the disappearance of the event horizon.
The following constraint must be satisfied: $ a^2 + Q^2 \le M^2.$

When the condition is violated, the event horizon disappears and
the solution describes a naked singularity instead of a black hole. Such odd
things should not exist in the real universe (this is the statement of the so--called
Cosmic Censorship Conjecture, not yet rigorously proved). For instance, for uncharged rotating
configuration, the condition $J_{max} = M^2$ corresponds to the vanishing of surface gravity
on the event horizon, due to ``centrifugal forces"; the corresponding solution is called extremal Kerr Solution.
Also, the maximal allowable electric charge is 
$Q_{max} = M \approx 10^{40} e \, M/M_{\odot}$, where $e$ is the electron charge, but it is to be noticed that
in realistic situations, black holes should not be significantly charged. This is due to the
extreme weakness of gravitational interaction compared to electromagnetic interaction.
Suppose  a black hole forms with initial positive charge $Q_i$ of order $M$. In realistic
conditions, the black hole is not isolated in empty space but is surrounded by charged
particles of the interstellar medium, e.g. protons and electrons. The black hole will
predominantly  attract electrons and repel protons with charge $e$ by its electromagnetic
field, and predominantly attract protons of mass $m_p$ by its gravitational field.
The repulsive electrostatic force on protons is larger than the
gravitational pull by the factor
$eQ/m_pM \approx e/m_p \approx 10^{18}$.
Therefore, the black hole will neutralize itself almost instantaneously. As
a consequence, the Kerr solution, obtained in equation (\ref{luminet:KN}) by putting $Q=0$, can be
used for any astrophysical purpose involving black holes. It is also a good approximation to
the metric of a (not collapsed) rotating star at large distance, but it has not been matched
to any known solution that could represent the interior of a star.

The Kerr metric in Boyer-Lindquist coordinates has singularities on the axis
of symmetry $\theta = 0$ -- obviously a coordinate singularity -- and for $\Delta
= 0$. One can write $\Delta = (r-r_+)(r-r_-)$ with $r_+ = M +\sqrt{M^2 -
a^2}$. The distance $r_+$ defines the outer event horizon (the surface of the rotating
black hole), whereas $r_-$ defines the inner event horizon. 
Like in Schwarzschild metric (where $r_+$ and $r_-$ coincide at the value $2M$),
the singularities at $r=r_+$, $r=r_-$ are coordinate singularities which can be
removed by a suitable transformation analogous to the ingoing
Eddington-Finkelstein coordinates for Schwarzschild space--time. For full mathematical
developments of Kerr black holes, see Chandrasekhar (1992) and O'Neill (1995).
 
\subsection {The black hole maelstrom}

There is a deep analogy between a rotating black
hole and the familiar phenomenon of a vortex - such as a giant
maelstrom produced by sea currents. If we cut a light cone at fixed time (a
horizontal plane in figure 8), the resulting spatial section is a
``navigation ellipse" which determines the limits of the permitted
trajectories. If the cone tips over sufficiently in the gravitational
field, the navigation ellipse detaches itself from the point of emission. The
permitted trajectories are confined within the angle formed by the
tangents of the circle, and it is impossible to go backwards. 
    
\begin{figure}[tb]
\renewcommand \thefigure {8\&9}
  \begin{center}
    \leavevmode
    \includegraphics{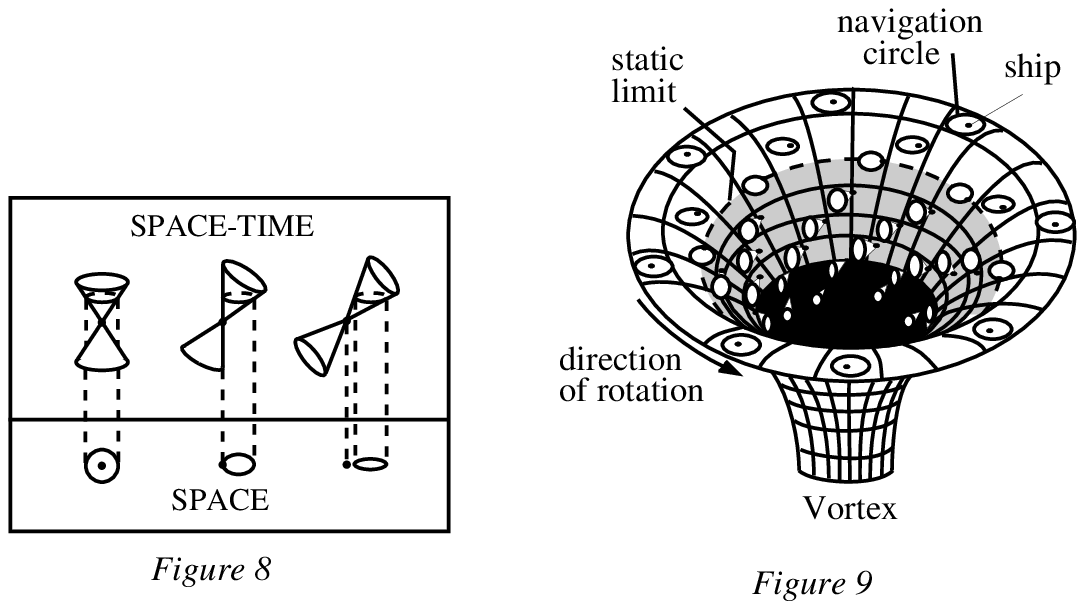}
    \caption{Navigation circles in the black hole maelstrom.}
  \end{center}
\end{figure}
\setcounter{figure}{9}
This projection
technique is useful to depict the causal structure of space-time around a
rotating black hole (figure 9). The gravitational well caused by a rotating
black hole resembles a cosmic maelstrom. A spaceship travelling in the vicinity is
sucked towards the centre of the vortex like a boat. In the region outside
the so-called {\it static limit} (clear), it can navigate 
to whereever it wants.
In the zone (in grey) comprised between the static limit and the event
horizon, it
is forced to rotate in the same direction as the black hole; its ability to
navigate freely is decreased as it is sucked inwards, but it can still escape
by travelling in an outwards spiral. The dark zone represents the region inside
the event horizon: any ship which ventured there would be unable to escape even
if it was travelling at the speed of light. A fair illustration is the 
Edgar Poe's short
story: {\it A descent into the maelstrom} (1840).

The static limit is a hypersurface of revolution,
given by the equation $r = M + \sqrt{M^2 - a^2 cos^2\theta}$. As we
can see from figure 10, it intersects the event horizon at
its poles $\theta = 0,\pi$ but it lies outside the horizon for other values
of $\theta$. The region between the static limit and the event horizon is
called ergoregion. There, all stationary observers must orbit the black hole
with positive angular velocity. The ergoregion contains orbits with negative
energy. Such a property has lead to the idea of energy extraction from a
rotating black hole.  Roger Penrose (1969) suggested the following mechanism.
A distant experimentalist
fires a projectile in the direction of the ergosphere along
a suitable trajectory (figure 10). When it arrives the projectile splits into two
pieces: one of them is captured by the black hole along a retrograde orbit,
while the other flies out of the ergosphere and is recovered by the
experimentalist.
Penrose has demonstrated that the experimentalist could direct the
projectile in such a way that the returning piece has a greater energy than
that of the initial projectile. This is possible if the fragment captured by the
black hole is travelling in a suitable retrograde orbit (that is orbiting in
the opposite sense to the rotation of the black hole), so that when it
penetrates the black hole it slightly reduces the hole's angular momentum. The
net result is that the black hole looses some of its rotational energy and the
difference is carried away by the escaping fragment. 

\begin{figure}[tb]
  \begin{center}
    \leavevmode
    \includegraphics{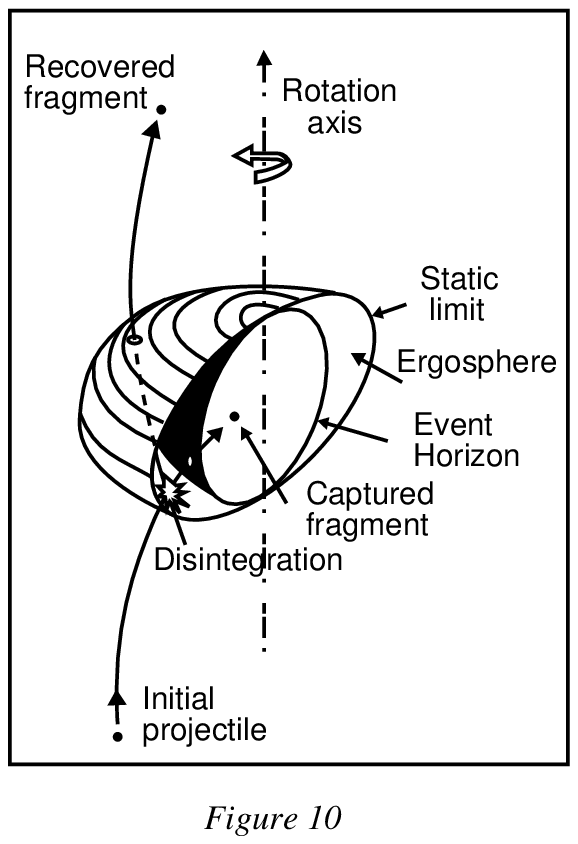}       
    \caption{Cross-section of a rotating black hole.}
  \end{center}
\end{figure}
The amount of energy that
can theoretically be extracted from a black hole has been calculated by
Christodolou and Ruffini (1971). The total mass-energy of a black hole is
\begin{equation}
        M^2 = \frac{J^2}{4M_{ir}^2} + (\frac{Q^2}{4 M_{ir}} + M_{ir})^2
\end{equation}
where 
$M_{ir} \equiv \frac{1}{2}\sqrt{\left(M+\sqrt{M^2-Q^2-a^2}\right)^2+a^2}$.
The first term corresponds to the rotational energy, the second one to the
Coulomb energy, the third one to an ``irreducible" energy.  The rotational
energy and the Coulomb energy are extractable by physical means such as
the Penrose process, the superradiance (analogous to stimulated emission in
atomic physics) or electrodynamical processes (see Norbert Straumann's
lectures for details), while the irreducible part
cannot be lowered by classical (e.g. non quantum) processes. The maximum
extractable  energy is as high a 29 per cent for rotational energy and 50 per
cent for Coulomb energy. It is much more efficient that, for instance,  nuclear
energy release (0.7 per cent for hydrogen burning). 

\subsection {Black hole thermodynamics}
\index{black hole!thermodynamics}
\label{sec:luminet:BHTD}

It is interesting to mention that the
irreducible mass is related to the area $A$ of the event horizon by 
$M_{ir} = \sqrt{A/16\pi}$.
Therefore the area of an event horizon cannot decrease
in time by any classical process. This was first noticed by Stephen
Hawking, who drew the striking analogy with ordinary thermodynamics, in which
the entropy of a system never decreases in time. Such a property has motivated a great deal
of theoretical efforts in the
1970's to better understand the laws of black hole dynamics -- i.e. the laws
giving the infinitesimal variations of mass, area and other black hole 
quantities when a black
hole interacts with the external universe -- and to push the analogy with thermodynamical
laws. For the development of black hole thermodynamics, see G. Neugebaueur's and W.
Israel's lectures in this volume. Let me just recall that black hole mechanics is
governed by four laws which mimic classical thermodynamics:
\begin{itemize}
\item {\it Zeroth law}.

In thermodynamics: all parts of a system at thermodynamical equilibrium have
equal temperature $T$.

In black hole mechanics: all parts of the event horizon of a black hole at
equilibrium have the same surface gravity $g$. The
surface gravity is given by the Smarr's formula $M = gA/4\pi + 2
\Omega_HJ + \Phi_H Q$, where  $\Omega_H$ is the angular velocity at the
horizon and  $\Phi_H$ is the co-rotating electric potential on the horizon.
This is a quite remarkable property when one compares to ordinary astronomical
bodies, for which the surface gravity depends on the latitude. Whatever a black
hole is flattened by centrifugal forces, the surface gravity is the same at
every point.
\item {\it First Law}.

In thermodynamics: the infinitesimal variation of the internal energy of a
system with temperature $T$ at pressure $P$ is related to the 
variation of entropy $dS$ and the variation of pressure $dP$ by
$dU = T dS - PdV$.

In black hole dynamics: the infinitesimal variations of the mass $M$, the charge
$Q$ and the angular momentum $J$ of a perturbed stationary black hole are
related by $dM = \frac{g}{8\pi}dA + \Omega_H dJ +\Phi_H dQ$.   
 
\item {\it Second Law}.

In thermodynamics, entropy can never decrease: $dS \ge 0$.

In black hole dynamics, the area of event horizon can never decrease: $dA
\ge 0$.

This law implies for instance that the area of a black hole resulting from
the coalescence of two parent black holes is greater than the sum of areas of
the two parent black holes (see Figure 11). It also implies that {\it black holes
cannot bifurcate}, namely a single black hole can never split in two parts. 

\item {\it Third law}.

In thermodynamics, it reflects the inaccessibility of the absolute zero of
temperature, namely it is impossible to reduce the temperature of a system to
zero by a finite number of processes.
  
In black hole mechanics, it is impossible to reduce the surface
gravity to zero by a finite number of operations. For Kerr black
holes, we have seen that zero surface gravity corresponds to the
 ``extremal" solution $J=M^2$.
\end{itemize}
 
\begin{figure}[tb]
  \begin{center}
    \leavevmode
    \includegraphics{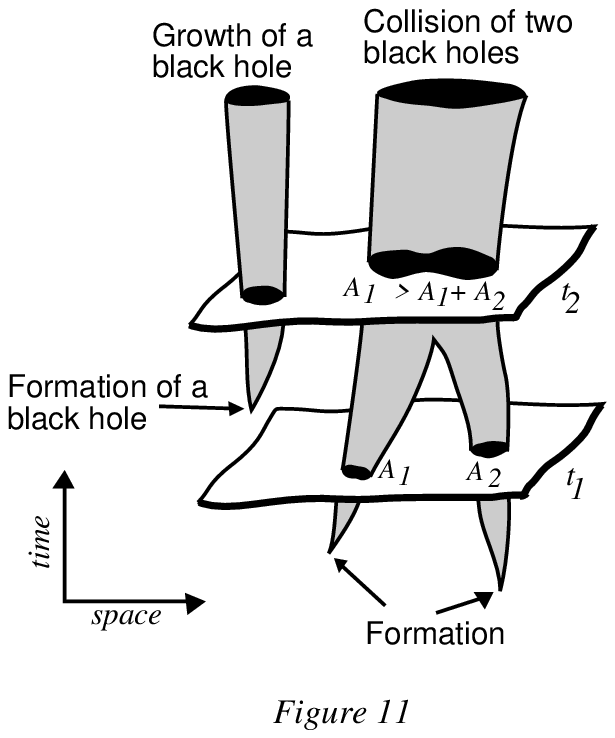}       
    \caption{The irreversible growth of black holes.}
  \end{center}
\end{figure}
It is clear that the area of the event horizon plays formally the
role of an entropy, while the surface gravity plays the role
of a temperature. However, as first pointed out by
Bekenstein, if black holes had a real temperature like thermodynamical systems,
they would radiate energy, contrarily to their basic definition. The puzzle
was solved by Hawking when he discovered the evaporation of mini-black
holes by quantum processes.

\subsection {The quantum black hole}
\index{quantum black hole}

The details of \index{Hawking radiation}Hawking radiation and the - not yet
solved - theoretical difficulties linked to its interpretation are discussed by other lecturers 
(Gerard 't Hooft, Andreas Wipf and Claus
Kiefer) in this volume. Therefore I shall only present the basic 
idea in a naive pictorial way (figure 12).
The black hole's gravitational field is described by (classical) general
relativity, while the surrounding vacuum space--time is described by quantum field theory.
The quantum evaporation process is analogous to pair production in a strong
magnetic field due to vacuum polarization. In the Fermi sea populated by
virtual pairs of particles-antiparticles which create and annihilate themselves,
the four various possible processes are depicted schematically in figure 
12.

\begin{figure}[tb]
  \begin{center}
    \leavevmode
    \includegraphics{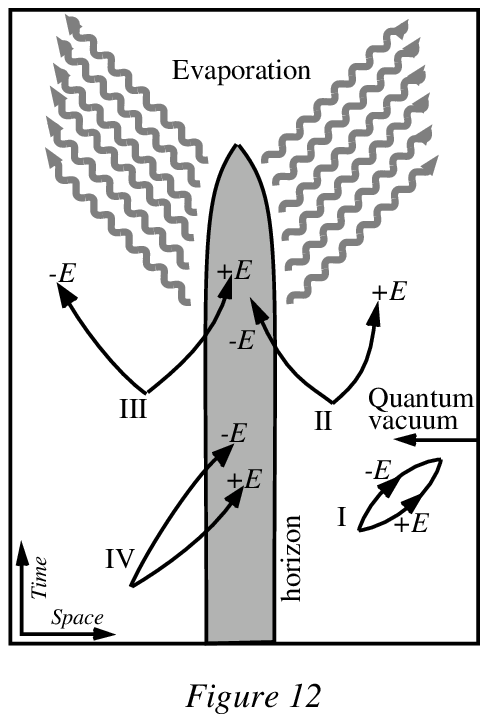}
    \caption{The quantum evaporation of a mini black hole by
      polarisation of the vacuum.}
  \end{center}
\end{figure}
Some virtual pairs  emerging from the quantum vacuum
just annihilate outside the horizon (process I). Some pairs produced in the
vicinity of the black hole disappear completely in the event horizon (process
IV). Some pairs are splitted, one particle (or antiparticle) escaping the black
hole, the other one being captured (processes II and III). The calculations show
that the process II is dominant, due to the (classical) gravitational 
potential which polarizes the quantum vacuum. As a consequence,
a black hole radiates particles
with a thermal spectrum characterized by a blackbody temperature precisely
given by the formula suggested by the thermodynamical analogy:
 \begin{equation}
  T = \hbar \,\frac{g}{2\pi} = 10^{-7} \frac{M_{\odot}}{M} \textrm{K},   
        \label{luminet:11}
 \end{equation} 
where  $\hbar$
is Planck's constant. We immediately see that the temperature is completely
negligible for any astrophysical black hole with mass comparable or greater 
to the solar mass. But for mini-black holes with mass $10^{15}$g (the typical
mass of an asteroid), the Hawking temperature is $10^{12} K$. 
Since the black hole  radiates away, it looses energy and evaporates on a
timescale approximately given by 
\begin{equation}
  t_E \approx 10^{10} years \, \left(\frac{M}{10^{15} grams}\right)^3      
        \label{luminet:12}
 \end{equation}
Thus, mini-black holes whose mass is smaller than that of an asteroid (and
 size less than $10^{-13} cm$) evaporate on a timescale shorter than the
age of the universe. Some of them should evaporate now and give rise to a huge
burst of high energy radiation. Nothing similar has ever been observed 
($\gamma$-ray bursts are explained quite differently). Such an observational
constraint thus limits the density of mini-black holes to be less than about
$100 /(lightyear)^3$. 

The black hole entropy is given by
\begin{equation}
        S = \frac{k_B}{\hbar} \frac{A}{4}
        \label{luminet:13}
\end{equation}
 (where $k_B$ is Boltzmann's constant), a formula which numerically
gives $S \approx 10^{77} k_B(\frac{M}{M_{\odot}})^2$ for a Schwarzschild black hole.
Since the entropy of a non--collapsed star like the Sun is approximately
$10^{58} k_B$,
we recover the deep meaning of the ``no hair'' theorem, according to
which black holes are huge entropy reservoirs.
By Hawking radiation, the irreducible mass, or equivalently the event 
horizon area of a black hole decreases, in violation of the Second Law of black hole mechanics. The
latter has to be generalized to include the entropy of matter in exterior
space--time. Then, the total entropy of the radiating black hole is
$S = S_{BH} + S_{ext}$ and, since the Hawking radiation is thermal, $S_{ext}$
increases, so that eventually $S$ is always a non-decreasing function of time.  

To conclude briefly the subject, even if mini-black holes are exceedingly rare, or even if they do not
exist at all in the real universe because the big bang could not have produced 
such fluctuations, they
represent a major theoretical advance towards a better understanding 
of the link between gravity
and quantum theory. 

\subsection{Space-time mappings}

Various mathematical techniques allow the geometer to properly visualize 
the complex space-time structure generated by black holes. 

\subsubsection{Embedding diagram --}

The space-time generated by a spherical mass $M$ has the 
Schwarzschild metric:
\begin{equation}
ds^2 = - \left(1- \frac{2M(r)}{r}\right)dt^2
+ \left(1-\frac{2M(r)}{r}\right)^{-1}dr^2 + r^2 d\Omega^2        
\end{equation}
where $M(r)$ is the mass comprised within the radius $r$. Since the 
geometry is static and spherically symmetric, we do not loose much 
information in considering only equatorial slices $\theta = \pi/2$ 
and time slices $t = constant$. We get then a curved 2--geometry with 
metric $(1-\frac{2M(r)}{r})^{-1}dr^2 + r^2 d\phi^2$. Such a surface 
can be visualized by embedding it in Euclidean 3--space
$ds^{2} = dz^{2} + dr^{2} + 
d\phi^{2}$. For a non--collapsed star with radius $R$, the outer solution
$z(r) = \sqrt{8M(r-2M)}$ for $r\geq R 
\ge 2M$ is asymptotically flat and matches exactly the non--singular inner solution 
$z(r) = \sqrt{8M(r)(r-2M(r))}$ for $0 \le r\le R$ (figure 13). For a 
black hole, the embedding is defined only for $r \geq 2M$. The 
corresponding surface is the Flamm paraboloid  $z(r) = 
\sqrt{8M(r-2M)}$. Such an asymptotically flat surface exhibits two 
sheets separated by the ``Schwarzschild throat" of radius $2M$ . The two sheets can be either considered as two 
different asymptotically flat ``parallel" universes (whatever the physical meaning of such 
a statement may be) in which a black hole in the upper sheet is 
connected 
to a time--reversed ``white hole" in the lower sheet (figure 14), or 
as a single asymptotically flat space--time containing a pair of 
black/white holes connected by a so--called ``wormhole" (figure 15). The freedom 
comes from the topological indeterminacy of general relativity, 
which allows us to identify some asymptotically distant points of 
space--time without changing the metric. 

\begin{figure}[htb]
\renewcommand \thefigure {13\&14}
  \begin{center}
    \leavevmode
    \includegraphics{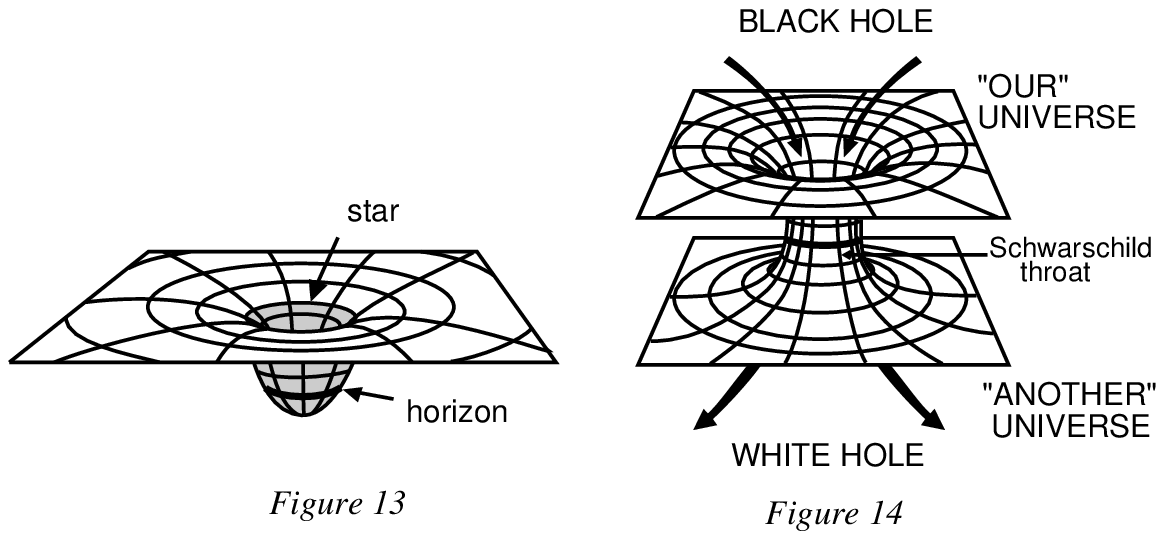}
    \caption{Embedding of a non-collapsed spherical star (13) and of
      Schwarzschild space-time (14).}
  \end{center}
\end{figure}
\setcounter{figure}{14}
\begin{figure}[htb]
  \begin{center}
    \leavevmode
    \includegraphics{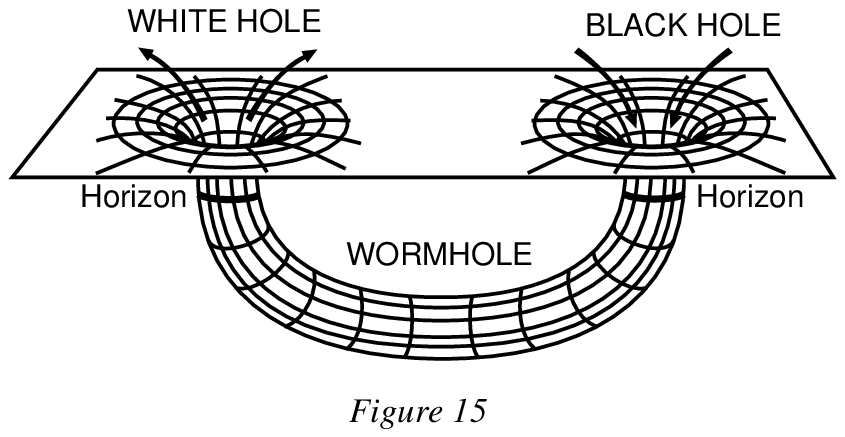}      
    \caption{A wormhole in space-time.}
  \end{center}
\end{figure}
However the embedding technique does not give access to the regions of 
space-time inside the event horizon.  

\subsubsection{Kruskal diagram --}
\index{Kruskal diagram}

To explore inner space-time we use the maximal analytical extension 
of the Schwarzschild metric. This is achieved by means of a coordinate 
transformation discovered by Kruskal: 
\begin{eqnarray}
\lefteqn{u^{2} - v^{2} = (\frac{r}{2M}-1)e^{r/2M}}\nonumber\\
 & 
\frac{v}{u} = 
\left\{\begin{array}{c}\coth {t\over4M}\\1\\\tanh {t\over4M}\end{array}\right\}
\quad\textrm{for}\quad
r\left\{\begin{array}{c}<2M\\=2M\\>2M\end{array}\right.
\end{eqnarray}
The metric then becomes

\begin{equation}
        ds^{2} = \frac{32M^3}{r} e^{-r/2M}(-dv^{2} + du^{2})+ 
        r^{2}d\Omega^{2}
        \label{luminet:14}
\end{equation}

In the $(v,u)$ plane, the Kruskal space-time divides into two outer 
asymptotically flat regions and two regions inside the event horizon 
bounded by the future and the past singularities. Only the unshaded 
region is covered by Schwarzschild coordinates. The black region does 
not belong to space-time. In the Kruskal 
diagram (figure 16), light rays  always travel at $45^{\circ}$, lines of 
constant distance $r$ are hyperbolas, lines of constant time $t$ pass 
through the origin. The interior of the 
future event horizon is the black hole, the interior of the past event 
horizon is the white hole. However it is clear that the wormhole 
cannot be crossed by timelike trajectories: no trajectory 
can pass from one exterior universe to the other one without 
encountering the $r=0$ singularity. 

\begin{figure}[tb]
  \begin{center}
    \leavevmode
    \includegraphics{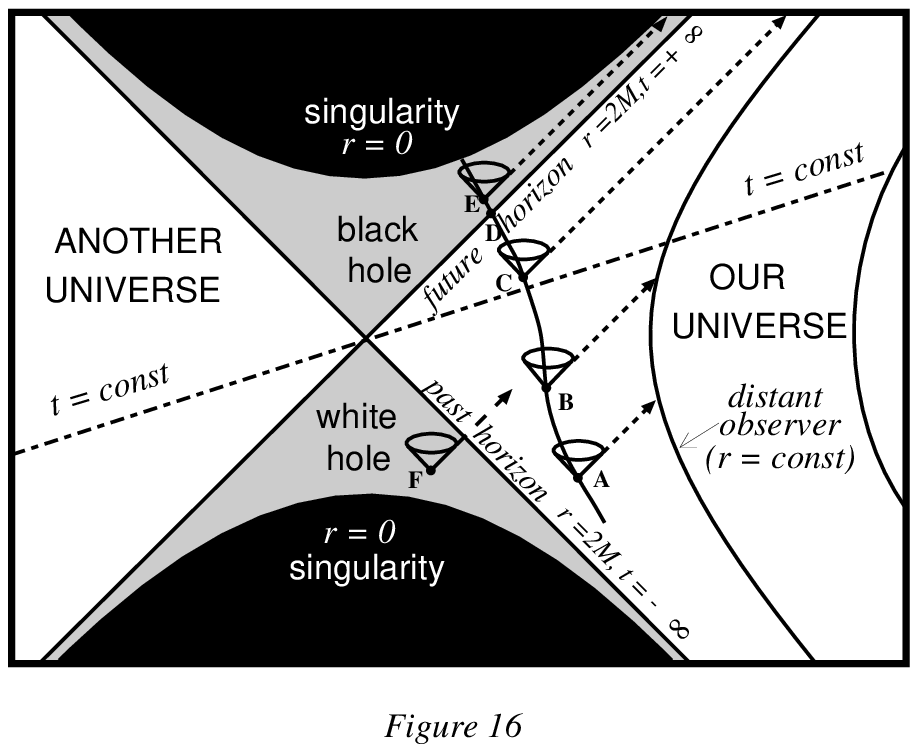}    
    \caption{Exploring a spherical black hole using Kruskal's map.}
  \end{center}
\end{figure}
\begin{figure}[tb]
  \begin{center}
    \leavevmode
    \includegraphics{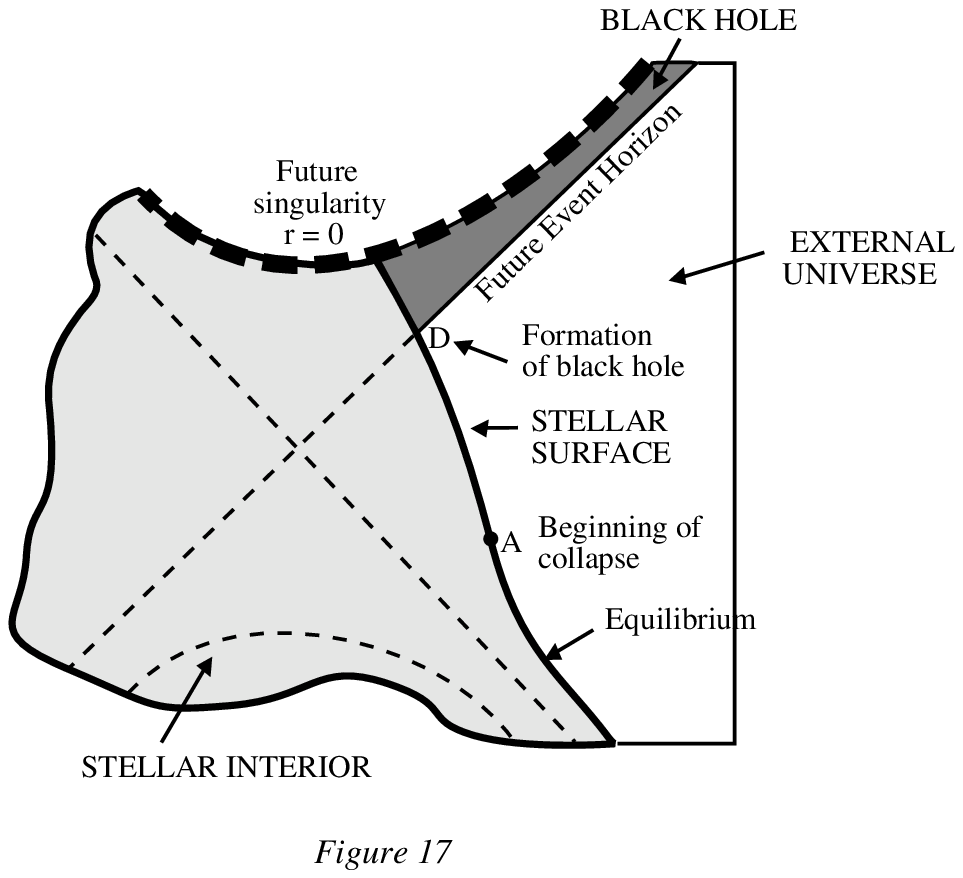}    
    \caption{Truncated Kruskal map representing the collapse of a star
      into a black hole.}
  \end{center}
\end{figure}
Moreover, the Kruskal extension is a mathematical idealization of a 
spherical black hole since it implicitly assumes that the black hole 
exists forever. However in the physical universe, a black hole is not inscribed
 in the initial conditions of the universe, it may form only from gravitational collapse. 
In such a case, one gets a ``truncated" Kruskal diagram (figure 17), 
which indicates that only the future event horizon and the future 
singularity occur in a single asymptotically flat space--time. Such 
a situation offers no perspective to space--time travelers!

\subsubsection{Penrose--Carter diagrams --}

The Penrose--Carter diagrams use conformal transformations of 
coordinates $g_{\alpha \beta} \to \Omega^{2}g_{\alpha \beta}$ which 
put spacelike and timelike infinities at finite distance, and thus 
allow to depict the full space--time into square boxes. The 
Penrose--Carter diagram for the Schwarzschild black hole does not 
bring much more information than the Kruskal one, but it turns 
out to be the best available tool to reveal the complex structure of a rotating black hole.
Figure 18 shows the ``many--fingered" universe of a Kerr black hole; 
it suggests that some timelike trajectories ($B,C$) may well cross the 
outer $EH$
and inner $IH$
event horizons and pass from an asymptotically flat external universe to 
another one without encountering a singularity. This is due to the fact 
that the singularity $S$ is timelike 
rather than spacelike. Also, the shape of the singularity is a ring 
within the equatorial plane, so that some trajectories ($A$) can pass 
through the ring and reach an asymptotically flat space--time inside 
the black hole where gravity is repulsive.
However, the analysis of perturbations of such idealized Kerr 
space--times suggests that they are unstable and therefore not physically plausible. 
Nevertheless the study of the  internal structure of black holes is a 
fascinating subject which is more deeply investigated in this 
volume by Werner Israel's lectures.

\begin{figure}[tb]
  \begin{center}
    \leavevmode
    \includegraphics{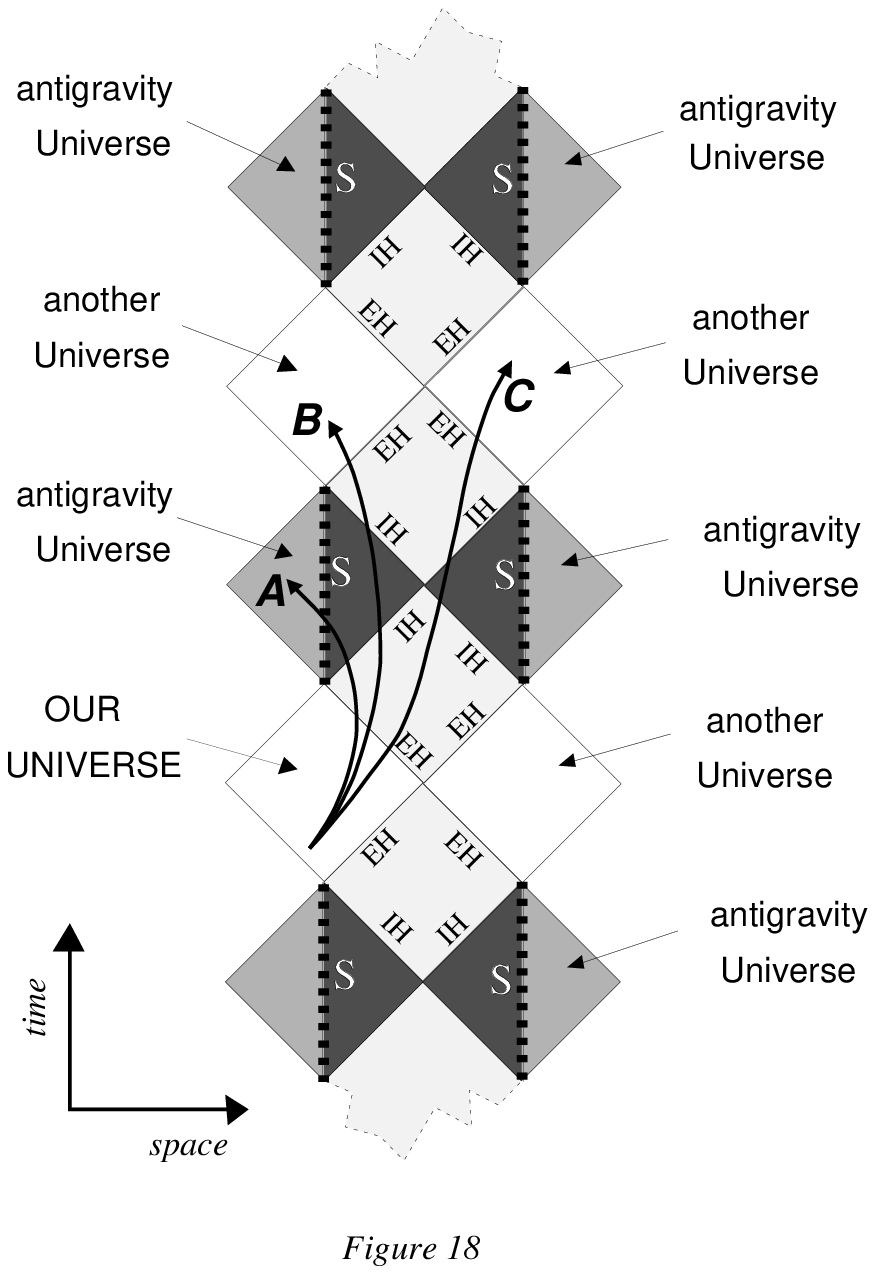}
    \caption{Penrose map of a rotating black hole.}
  \end{center}
\end{figure}

\section{Astrophysics of Black Holes}

The fact that General Relativity does predict the existence of black holes and that
General Relativity is a reliable theory of gravitation does not necessarily prove the
existence of black holes, because General Relativity does not describe the
astrophysical processes by which a black hole may form.

Thus, the astronomical credibility of black holes crucially depends on a good
understanding of gravitational collapse of stars and stellar clusters.

In this section we first examine briefly the astrophysical conditions for 
black hole formation, next we describe the astronomical sites where 
black hole candidates at various mass scales lurk.

\subsection {Formation of stellar black holes}

The basic process of stellar evolution is gravitational contraction at a rate
controlled by luminosity. The key parameter is the initial mass. According to its
value, stars evolve through various stages of nuclear burning and 
finish their lives as white
dwarfs, neutron stars or black holes. Any stellar remnant (cold equilibrium configuration)
more massive than about $3 M_{\odot}$ cannot be supported by degeneracy pressure and
is doomed to collapse to a black hole. 

\begin{figure}[tb]
  \begin{center}
    \leavevmode
    \includegraphics{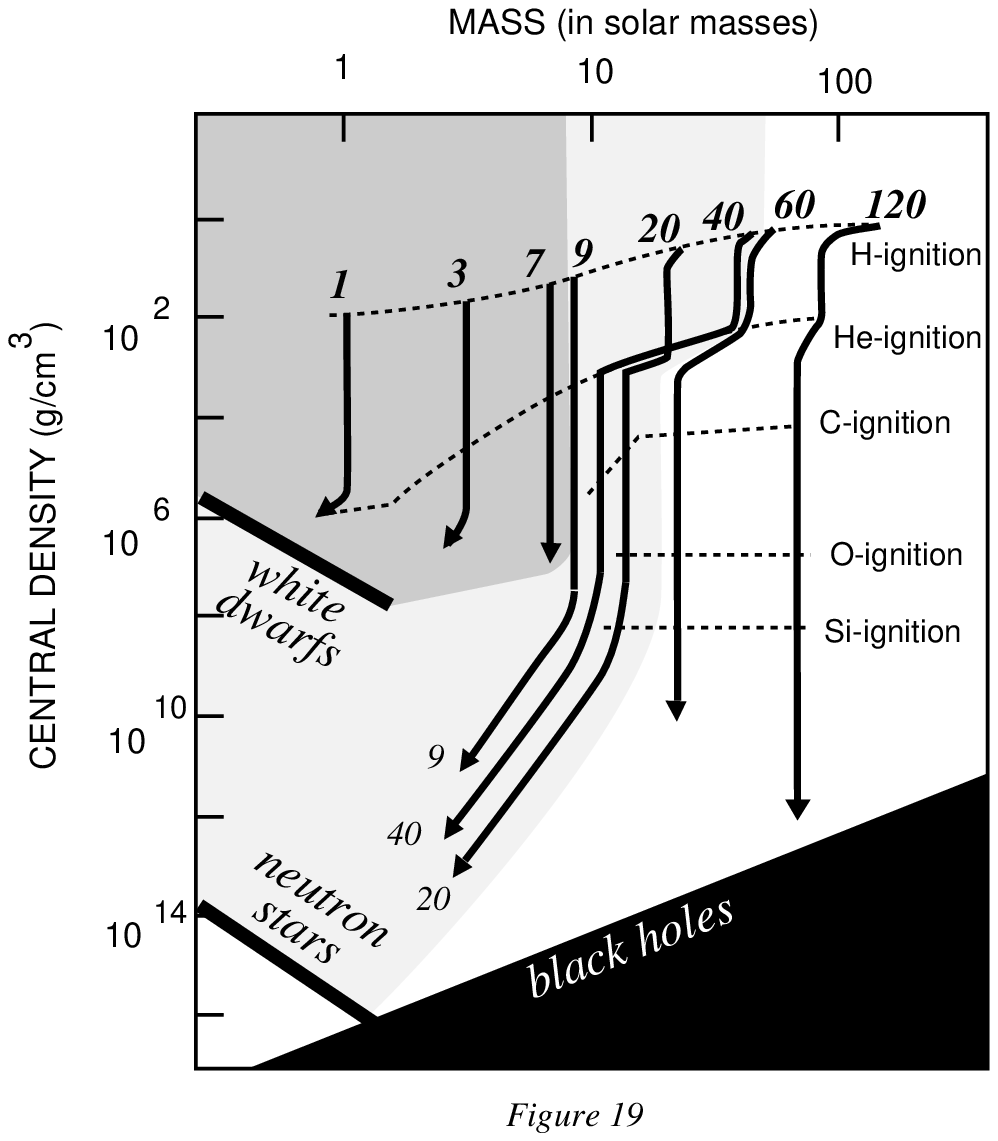}
    \caption{The density-mass diagram of astronomical objects.}
  \end{center}
\end{figure}
The figure 19 shows the stellar paths in a
density-mass diagram according to the most recent observational and 
theoretical data. Below $8 M_{\odot}$ stars produce white dwarfs, between 8 and $45
M_{\odot}$ they produce neutron stars; black holes are formed only 
when the initial mass
exceeds $45 M_{\odot}$ (we note on the diagram that stars with 
initial mass between 20 and $40 M_{\odot}$ suffer important mass 
losses at the stage of helium burning). 
Taking account of the stellar initial mass function, one concludes that 
approximately 1 supernova
over 100 generates a black hole rather than a neutron star.
Another possibility to form a stellar mass black hole is
accretion of gas onto a neutron star in a binary system until when the mass of
the neutron star reaches the maximum allowable value; then, gravitational 
collapse
occurs and a low mass black hole forms. 
Taking into account these various processes, a typical galaxy like the Milky 
Way should harbour $10^7 - 10^8$ stellar black holes.

\subsection {Formation of giant black holes}

A massive black hole can form by continuous growth of a ``seed" stellar mass black
hole, by gravitational collapse of a large star cluster or by collapse 
of a large density
fluctuation in the early universe (see next subsection).
A well nourished stellar mass black hole can grow to a
supermassive black hole in less than a Hubble time. Such a process requests large
amounts of matter (gas and stars) in the neighborhood, a situation
than can be expected in some galactic nuclei.

A dense cluster of ordinary stars, such that the velocity dispersion $v_c \leq
v_*$, where $v_* \approx 600 \, km/s$ is the typical escape velocity for main sequence
stars, first evolves through individual stellar burning;
supernovae explosions lead to the formation of compact remnants, e.g. neutron stars
and stellar mass black holes. A cluster of compact stars becomes relativistically unstable at
sufficiently high central gravitational redshift $1+z_c = (1-2M)^{-1/2} \geq 1.5$
(Zeldovich and Podurets, 1965). Numerical simulations (e.g. Shapiro and Teukolsky, 1987,
Bisnovatyi-Kogan, 1988) confirm this scenario. Starting with $\approx 10^7 - 2.10^8$ compact
stars of $1-10 M_{\odot}$ within a cluster radius $r \leq 0.01 - 0.1 \,$pc and velocity
dispersion $800-2000 \, $km/s, the evolution proceeds through three stages: 
\begin{itemize}
\item secular core collapse via the gravothermal catastrophe (long timescale)
\item short epoch dominated by compact star collisions and coalescences, leading to the
formation of black holes with mass $M \approx 90 M_{\odot}$
\item relativistic instability leading to a massive black hole surrounded by a halo of
stars.
\end{itemize}    
\subsection {Formation of mini black holes}

Zeldovich in 1967 and Hawking in 1971 pointed out that it was in principle 
possible to create a
black hole with small mass (e.g. below the Chandrasekhar limit) by applying 
a sufficiently strong external
pressure. Such conditions could have been achieved only in the very early universe. 
Gravitational forces may locally halt the cosmic expansion of a clump of 
matter and reverse it
into collapse if the self-gravitational potential energy of the clump exceeds the internal
energy:
\begin{equation}
\frac{GM^2}{R}  \approx G\rho ^2 R^5 \geq pR^3 \label{luminet:Jeans}
\end{equation}
 During the radiation era, $p \approx
\rho c^2$, so the condition (\ref{luminet:Jeans}) is equivalent 
to $GM/c^2 \geq R$, where $R$ is the size of the fluctuation. Then a primordial 
black hole of mass $M$
forms. Due to the relation between density and time $G\rho \approx t^{-2}$ in 
an Einstein-De Sitter model 
of the early universe, the maximum mass of a
collapsing fluctuation is related to the cosmic time by $M(grams) \approx 
10^{38} t \,(seconds)$. 
Thus at Planck time $t \approx 10^{-43}$s, only mini black holes may form with the Planck mass
$\approx 10^{-5}$g, at time $t  \approx 10^{-4}$s, black holes may form with
$\approx 1 M_{\odot}$, at the time of nucleosynthesis $t \approx 100 
\,$s supermassive
black holes with $10^7 M_{\odot}$ may form.
The observational status of primordial black holes is poor and 
unclear. On one hand, mini black holes with mass $\leq 10^{15}$g could be detected by a burst of
$\gamma$--radiation corresponding to the last stage of quantum evaporation in less than a
Hubble time. Nothing similar having been observed, this puts severe upper limits on the actual
average density of mini-black holes. On the other hand, the fact that most 
galactic nuclei seem to harbour massive black holes (see below) and that 
supermassive black holes are suspected to feed
quasars at very high redshift, favour the hypothesis of the rapid formation 
of primordial massive black holes in
the early universe.

\subsection {Black Hole candidates in binary X-ray sources}

Light cannot escape (classical) black holes but one can hope to detect 
them indirectly by observing the electromagnetic energy released 
during accretion processes.

Accretion of gas onto a compact star (neutron star or black hole in 
a binary system) releases energy in the X-ray domain, see S. 
Chakrabarti's and J. Tr\"umper's lectures in this volume for the details. 
Search for stellar mass black holes thus consists in locating rapidly 
variable binary 
X-ray sources which are neither periodic (the corresponding {\it X-ray 
pulsars} 
are interpreted as rotating neutron stars) nor recurrent (the 
corresponding {\it X-ray bursters} are interpreted as thermonuclear 
explosions on a neutron star's hard surface). In spectroscopic 
binaries, the Doppler curve of the spectrum of the primary (visible) star 
provides the orbital period $P$ of the binary and the maximum 
velocity $v_{*}$ of 
the primary projected along the line-of-sight. Kepler's laws 
gives the following mass function which relates observed quantities to 
unknown masses: 
\begin{equation}
        \frac{Pv_{*}^3}{2 \pi G} = \frac{(M_{c}\sin i)^{3}}{(M_{*}+M_{c})^{2}}
        \label{luminet:15}
\end{equation}
where $M_{c}$ and $M_{*}$ are the masses of the compact star and of the 
optical primary, $i$ the orbital inclination angle.
A crucial fact is that $M_{c}$ cannot be less than the value of the 
mass function (the limit would correspond to a zero-mass companion 
viewed at maximum inclination angle). Therefore the best black hole 
candidates are obtained when the observed mass function exceeds 
$3 M_{\odot}$ -- since, according to the theory, a 
neutron star more massive than this limit is unstable and will 
collapse to form a black hole. Otherwise, additional information is 
necessary to deduce $M_{c}$: the spectral 
type of the primary gives approximately $M_{*}$, the presence or 
absence of X-ray eclipses gives bounds to $\sin i$. Hence $M_{c}$ is 
obtained within some error bar. Black hole candidates are retained 
only when the lower limit exceeds $3 M_{\odot}$. At present day, 
about ten binary X--ray 
sources provide good black hole candidates. They can be divided into 
two families: the high--mass X--ray binaries (HMXB), where the 
companion star is of high mass, and the low--mass X--ray binaries 
(LMXB) where the companion is typically below a solar mass. The latter 
are also called ``X--ray transients" because they flare up to high 
luminosities. Their mass properties are summarized in 
the table 1 below.

\begin{table}
\caption{\label{luminet:B}
{\bf Stellar mass black hole candidates}}
\begin{center}
\begin{tabular}{|c|c|c|c|}
\hline
      & mass function & $M_{c}/M_{\odot}$ & $M_{*}/M_{\odot}$     \\ \hline
{\it HMXB}  &      &  &      \\
            &      &  &      \\
 Cygnus X-1     & 0.25      &  11--21 &  24--42  \\
LMC X-3  & 2.3 & 5.6 --7.8 &  20  \\
LMC X-1      &0.14  & $\geq$ 4 &  4--8 \\ \hline
{\it LMXB (X--ray transients)}      &  &   &  \\
            &      &  &      \\
            V 404 Cyg      & 6.07  & 10--15 & $\approx$ 0.6  \\ 
A 0620-00      & 2.91  &  5--17 &  0.2--0.7 \\
GS 1124-68 (Nova Musc)      & 3.01  &  4.2--6.5 &  0.5--0.8  \\
GS 2000+25 (Nova Vul 88)  & 5.01 &  6-14 & $\approx$ 0.7  \\
   GRO J 1655-40   & 3.24& 4.5 -- 6.5 & $\approx$ 1.2\\
  H 1705-25 (Nova Oph 77)    & 4.65 &  5--9 &  $\approx$ 0.4\\
J 04224+32      & 1.21  & 6--14 & $\approx$ 0.3 -- 0.6  \\
\hline
\end{tabular}
\end{center}
\end{table}

Other galactic X--ray sources are suspected to be black holes on 
spectroscopic or other grounds, see Chakrabarti's lectures in this 
volume for developments. For instance, some people argue that gamma-ray 
emission (above $100 \, keV$) emitted from the inner edge of the 
accretion disc would attest the presence of a black hole rather than a 
neutron star, because the high-energy radiation is scattered back by the neutron star's hard surface and cools down the 
inner disc. If this is true, then many ``gamma--ray novae" in which no 
measurement of mass can be done (due to the absence of optical 
counterpart or other limitations) are also good black hole 
candidates. This is specially the case for Nova Aquila 1992 and 1 E 
17407-2942, two galactic sources which also exhibit radio jets. Such 
 ``microquasars" involving both accretion and ejection of matter provide an interesting 
 link between high energy phenomena at the stellar and galactic scales, 
 see J. Tr\"umper's lectures.

\subsection {Evidence for massive black holes in galactic nuclei}
\index{galactic nuclei}

After the original speculations of Michell and Laplace, the idea of 
giant black holes was reintroduced in the 1960's to explain the 
large amounts of energy released by {\it active galactic nuclei} (AGNs). This 
generic term covers a large family of galaxies including quasars, 
radiogalaxies, Seyfert galaxies, blazars and so on, for the 
classification  see J. Tr\"umper's, W. Collmar's and V. Sch\"onfelder's
lectures in this volume.
The basic process is accretion of gas onto a massive 
black hole. The maximum luminosity for a source of mass $M$, called 
the Eddington luminosity, is obtained by balance between 
gravitational attraction and radiation pressure repulsion acting on a 
given element of gas. It is given by 

\begin{equation}
L \approx 10^{39} \, \frac{M}{10^{8}M_{\odot}} \textrm{W}
        \label{luminet:16}
\end{equation}

The observed luminosities of AGNs range from $10^{37} -10^{41}$ 
W, where the higher values apply to the most powerful 
quasars.  Then the corresponding masses range from $10^{6} - 
10^{10} M_{\odot}$. 

Due to constant improvements of 
observational techniques, it turned out in the 1990's that most of the galactic 
nuclei (active or not)  harbour large mass concentrations. Today the 
detection of such masses is one of the major goals of  
extragalactic astronomy. The most convincing method of detection consists in 
the dynamical analysis of surrounding matter: gas or stars near 
the invisible central mass have large dispersion velocities, which can 
be measured by spectroscopy. It is now likely that giant black holes 
lurk in almost all galactic nuclei, the energy output 
being governed by the available amounts of gaseous fuel. The best 
candidates are summarized in Table 2. 

For instance, our Galactic Centre is observed in radio, infrared, X--ray and gamma--ray 
wavelengths (other wavelengths are absorbed by dust clouds of 
the galactic disc). A unusual radiosource has long been observed at 
the dynamical centre, which can be interpreted as low--level accretion 
onto a moderately massive black hole. However, a definite proof is not yet 
reached because gas motions are hard to interpret. Recently Eckart 
and Genzel (1996) obtained a full three--dimensional map of the 
stellar velocities within the central 0.1 pc of our Galaxy. The 
values and distribution of stellar velocities are convincingly 
consistent with the  hypothesis of a $2.5 \times 10^{6} M_{\odot}$ black 
hole.

The nucleus of the giant elliptical M87 in the nearby Virgo cluster has also a long 
story as a supermassive black hole candidate. Several independent 
observations are consistent with a $1-3 \, 10^{9} M_{\odot}$ black hole 
accreting in a slow, inefficient mode. A disc of gas is orbiting in a 
plane perpendicular to a spectacular jet; recent spectroscopic 
observations of the Hubble Space Telescope show redshifted and 
blueshifted components of the disc, which can be interpreted by Doppler effect as parts 
of the disc on each side of the hole are receding and approaching from us. 

The spiral galaxy NGC 4258 (M 106) is by far the best massive black hole 
candidate. Gas motions near the centre has been precisely mapped with 
the 1.3 cm maser emission line of $H_{2}O$. The velocities are 
measured with accuracy of 1 km/s. Their spatial distribution reveals a 
disc with rotational velocities following an exact Kepler's law around 
a massive compact object. Also the inner edge of the disc, orbiting at 
1080 km/s, cannot comprise a stable stellar cluster with the 
inferred mass of $3.6 \times 10^{7}M_{\odot}$.

 \begin{table}
\caption{\label{luminet:BB}
{\bf Massive black hole candidates}}
\begin{center}
\vspace{1.cm}
\begin{tabular}{|c|c|c|c|}
\hline
dynamics      & host galaxy & galaxy type & $M_{h}/M_{\odot}$     \\ \hline

maser     & M 106     &  barred            & $ 4 \times 10^{7}$  \\
gas       & M 87      &  elliptical        & $ 3 \times 10^{9}$ \\
gas       & M 84      & elliptical         & $ 3 \times 10^{8}$ \\
gas       & NGC 4261  & elliptical         & $ 5 \times 10^{8}$ \\           \hline
stars     & M 31      & spiral             & $3-10 \times10^{7}$  \\ 
stars     & M 32      & elliptical         & $ 3 \times 10^{6}$ \\
stars     & M 104     & (barred?) spiral   & $ 5-10 \times 10^{8}$  \\
stars     & NGC 3115  & lenticular         & $ 7-20 \times 10^{8}$\\
stars     & NGC 3377  & elliptical         & $ 8 \times 10^{7}$\\
stars     & NGC 3379  & elliptical         & $ 5 \times 10^{7}$\\
stars     & NGC 4486B & elliptical         & $ 5 \times 10^{8}$\\
stars     & Milky Way & spiral             & $ 2.5 \times 10^{6}$\\
\hline
\end{tabular}
\end{center}
\end{table}

The black hole in our galaxy and the massive black holes suspected in nearby 
ordinary galaxies would be small scale versions of the cataclysmic 
phenomena occurring in AGNs. But AGNs are too far away to offer a
spectral resolution good enough for dynamical measurements. Indeed, 
estimates of luminosities of AGNs and theoretical arguments involving 
the efficiency of energy release in strong gravitational fields 
invariably suggest that central dark masses are comprised between $10^{7} - 10^{9} \, M_{\odot}$. 
Variability of the flux on short timescales also 
indicates that the emitting region has a small size; many AGNs exhibit large
luminosity fluctuations over timescales as short as one hour, which indicate 
that the emitting region is smaller than one light-hour. Such large masses 
in such small volumes cannot be explained by star clusters, so that accreting massive
 black holes remain the only plausible explanation.

\subsection {Stellar disruption}
\index{Stellar disruption}

The accretion of gas at rate  $dM/dt$ and typical efficiency 
$\epsilon \approx 0.1$ 
produces a luminosity 
\begin{equation}
        L \approx 10^{39} (\frac{\epsilon}{0.1}) \, \frac{dM/dt}{1 
        M_{\odot}/year}\textrm{W}
\end{equation}  

By comparing the luminosity of the accretion model with the 
observed luminosities in AGNs, we conclude that the gas accretion rate 
must lie in the range $10^{-2}- 10^{2} \, M_{\odot}/year$. One is thus 
led to the question about the 
various gas production mechanisms able to fuel a giant black hole. An 
efficient process is mass loss from stars passing near the hole. 
Current models of galactic nuclei involve a massive black hole 
surrounded by a dense large cloud of stars. Diffusion of orbits makes 
some stars to penetrate deeply within the gravitational potential 
of the black hole along eccentric orbits. Disruption of stars can occur
either by \index{tidal force}tidal forces or 
by high-velocity interstellar collisions (figure 20). 
The collision radius $R_{coll} \approx 7 \times 10^{18} \,
\frac{M}{10^8 M_{\odot}}$cm
for a solar--type star is the distance within which the 
free--fall velocity of stars becomes greater than the escape velocity 
at the star's surface $v_{*}$ (typically 500 km/s for ordinary 
stars); if two stars collide inside $R_{coll}$ they will be 
partially or totally disrupted.

\begin{figure}[tb]
  \begin{center}
    \leavevmode
    \includegraphics{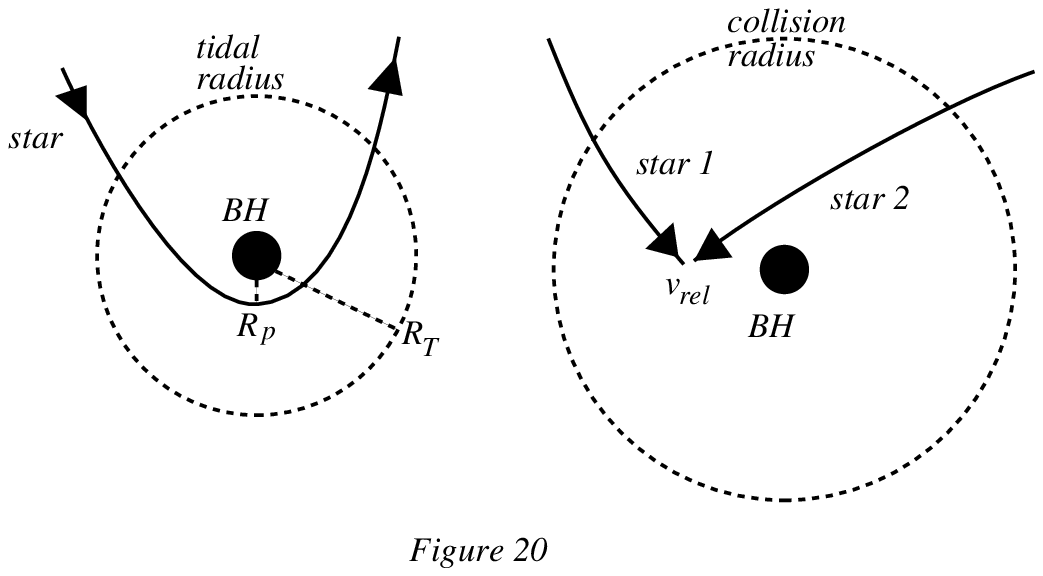}
    \caption{Tidal and collision radius.}
  \end{center}
\end{figure}
Also stars penetrating the critical \index{tidal radius}tidal radius
$R_{T} \approx 6\times
 10^{13} \, ( \frac{M}{10^{8}M_{\odot}} )^{1/3}$cm for solar--type 
stars will be ultimately disrupted, about 50 per cent of the released gas 
will remain bound to the black hole. In some sense, the tidal encounter 
of a star with a black hole can be considered as a collision of the 
star with itself\ldots 

In the collision process the factor $\beta = v_{rel}/v_{*}$ plays a 
role analogous to the penetration factor $\beta = 
R_{T}/R_{p}$ in the tidal case (where $R_{p}$ is the 
periastron distance). As soon 
as $\beta \geq 1$ the stars are disrupted, and when $\beta \geq 5$ the 
stars are strongly compressed during the encounter. Thus, in both cases,
$\beta$ appears 
as a {\it crushing factor}, whose magnitude dictates the degree of maximal
compression and heating of the star. 

The first modelisation of the tidal disruption of stars by a big 
black hole was done in the 1980's by myself and collaborators 
(see Luminet and Carter, 1986 and references therein). 
We have discovered that a star deeply 
plunging in the tidal radius without crossing the event horizon is 
squeezed by huge tidal forces and compressed into a short-lived, 
ultra-hot ``pancake" 
configuration. Figure 21 shows schematically the progressive 
deformation of the star (the size of the star has been considerably 
over-emphasized for clarity). The left diagram represents the deformation 
of the star in its orbital plane (seen from above), the right one shows 
the deformation in the perpendicular direction. From $a$ to $d$ the 
tidal forces are weak and the star remains practically spherical. At 
$e$ the 
star penetrates the tidal limit. It becomes cigar-shaped. From $e$ 
to $g$ a ``mangle" effect due to tidal forces becomes increasingly important and the 
star is flattened in its orbital plane to the shape of a curved 
``pancake". The star rebounds, and as it leaves the tidal radius, it 
starts to expand, becoming more cigar-shaped again. A little further 
along its orbit the star eventually breaks up into fragments.

If the star chances to penetrate deeply (says with $\beta \geq 10$), 
its central temperature increases to a billion degrees in a tenth of a 
second. The thermonuclear chain reactions are considerably enhanced. 
During this brief period of heating, elements like helium, nitrogen and 
oxygen are instantaneously transformed into heavier ones by rapid 
proton or alpha--captures. A thermonuclear explosion 
takes place in the stellar pancake, resulting in a kind of 
``accidental supernova". The consequences of such an explosion are far 
reaching. About 50 per cent of the stellar debris is blown away from 
the black hole at high velocity (propelled by thermonuclear energy 
release), as a hot cloud able of carrying away any other clouds it 
might collide with. The rest of the debris falls rapidly towards the 
hole, producing a burst of radiation. Like supernovae, the stellar 
pancakes are also crucibles in which heavy elements are produced and 
then scattered throughout the galaxy. Thus, 
observation of high--velocity clouds and enrichment of 
the interstellar medium by specific isotopes in the vicinity of galactic nuclei would 
constitute an observational signature of the presence of big black holes.

\begin{figure}[tb]
  \begin{center}
    \leavevmode
    \includegraphics{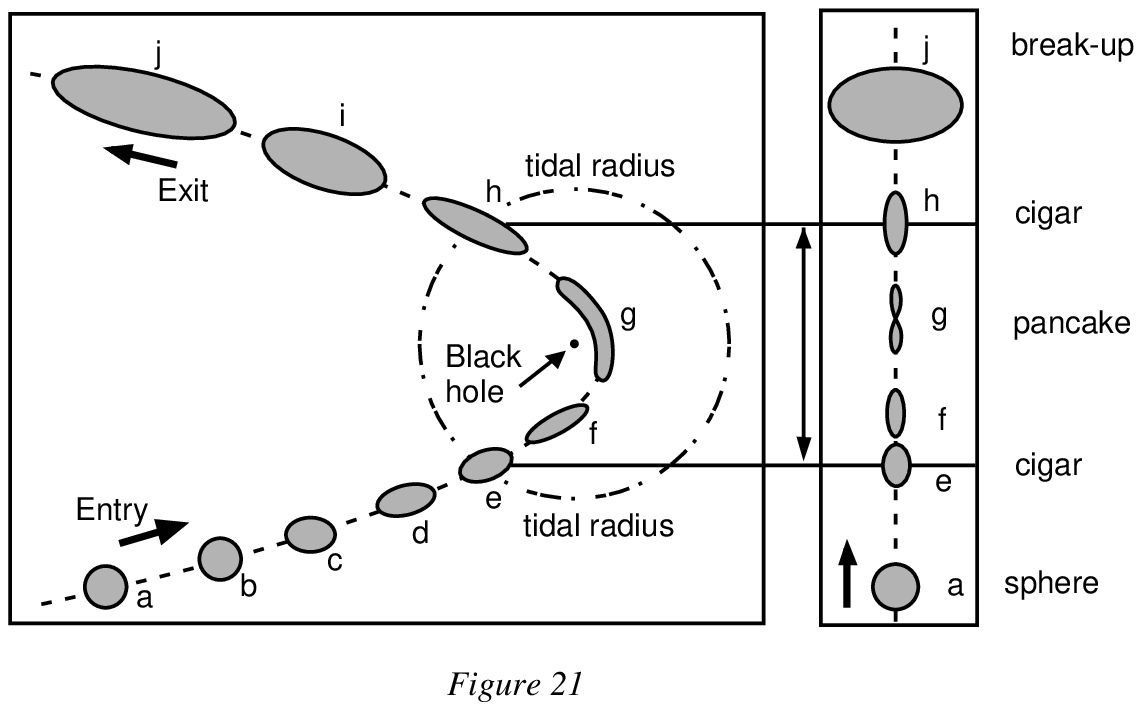}
\caption{The disruption of a star by tidal forces of a black hole.}
  \end{center}
\end{figure}
Explosive or 
not, the tidal disruption process would induce a burst of luminosity 
in the host galactic nucleus on a timescale of a few months (the time required 
for the debris to be digested).
To describe the evolution of the star we developed a simplified 
``affine model" in which we assumed that the layers of constant 
density keep an ellipsoidal form. Many astrophysicists were skeptical 
about the predictions of the model until when full hydrodynamical calculations were 
performed all around the world, using 3D Smooth Particle 
Hydrodynamical codes (Laguna and Miller, 1993, Khlokov, Novikov and 
Pethick, 1993, Frolov et al., 1994) or spectral methods (Marck et 
al., 1997). The main features and quantitative predictions of the 
affine star model were confirmed, even if shock waves may decrease a 
little bit the maximum pancake density.

The nucleus of the elliptical galaxy NGC 
4552 has increased its ultra-violet luminosity up to $10^{6} 
L_{\odot}$ between 1991 and 1993 (Renzini et al, 1993). The timescale 
was consistent with a tidal disruption process, however the luminosity 
was $\approx 10^{-4}$ lower than expected, suggesting only a partial disruption 
of the star.

\section {A journey into a black hole}

Imagine a black hole surrounded by a bright disc (Figure 22). The 
system is observed  from a great distance at an angle of $10^{\circ}$ 
above the plane of the disc. The light rays are received on a 
photographic plate (rather a bolometer in order to capture all 
wavelengths). Because of the curvature of space--time in the 
neighborhood of the black hole, the image of the system is very 
different from the ellipses which would be observed if an ordinary 
celestial body (like the planet Saturn) replaced the black hole. The 
light emitted from the upper side of the disc forms a direct image and 
is considerably distorted, so that it is completely visible. There is 
no hidden part. The lower side of the disc is also visible as an 
indirect image, caused by highly curved light rays. 
\begin{figure}[tb]
  \begin{center}
    \leavevmode
        \includegraphics{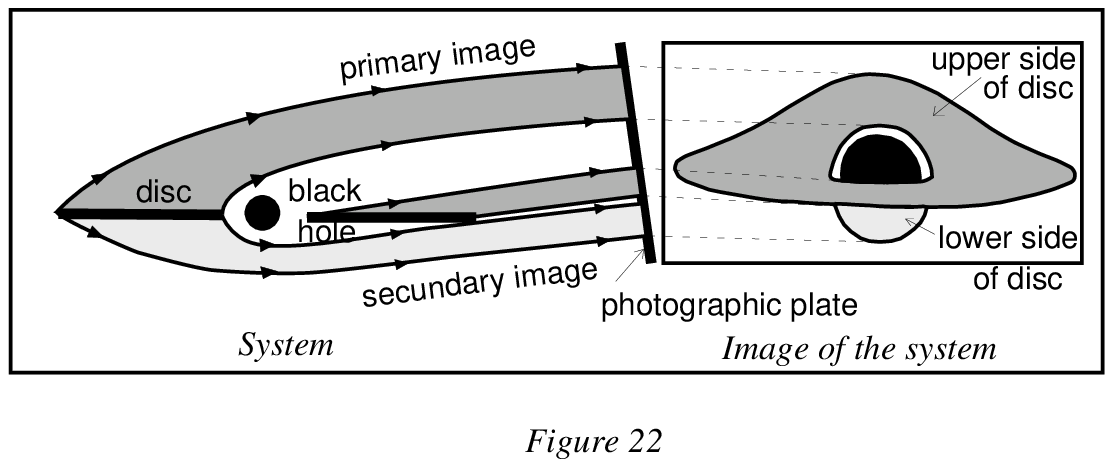}
        \caption{Optical distortions near a black hole.}
  \end{center}
\end{figure}

\begin{figure}[ht]
  \begin{center}
    \leavevmode
    \includegraphics{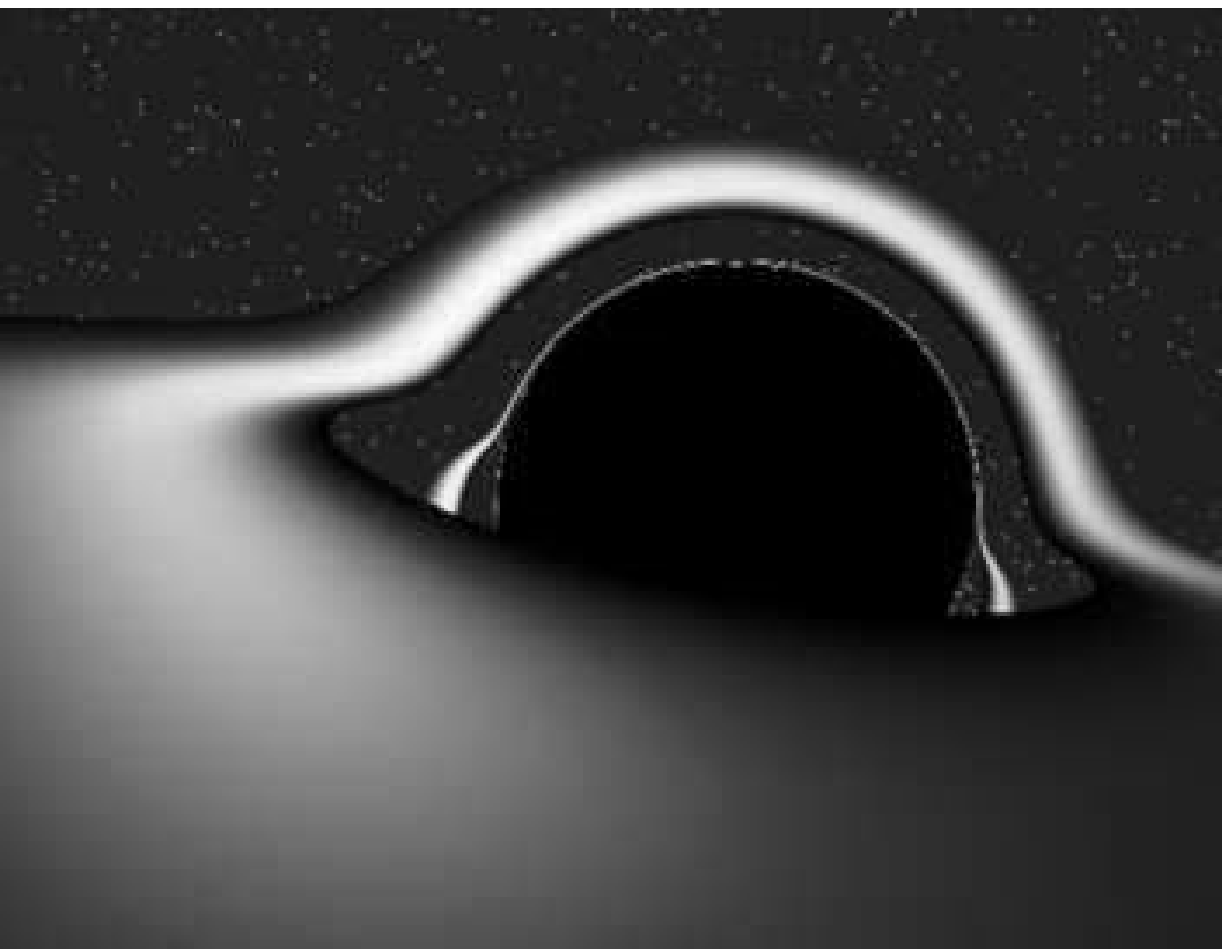}
    \caption{
The appearance of a distant black hole surrounded by an accretion disc.
(\copyright J.-A.Marck/Sygma)}
  \end{center}
\end{figure}
The first computer images of the appearance of a black hole surrounded 
by an accretion disc were obtained by myself (Luminet, 1978). More 
sophisticated calculations were performed by Marck (1993) in 
Schwarzschild and Kerr space--times. A realistic image, e.g. taking account 
of the space--time curvature, of the blue- and redshift effects, of the physical properties
of the disc and so on, can be precisely calculated 
at any point of space--time -- including inside the event horizon. A movie
showing the distortions observed along any 
timelike trajectory around a black hole was produced (Delesalle, 
Lachi\`eze-Rey and Luminet, 1993). The figure 23 is a snapshot taken along a parabolic 
plunging trajectory. During such a ``thought 
journey" the vision of the third butterfly becomes accessible, all external spectators 
can admire the fantastic landscape generated by the black hole. 

For a long time considered by astronomers as a mere theoretical speculation,
black holes are now
widely accepted as the basic explanation for X--ray massive binaries and galactic nuclei.
Allowing for the elaboration of the most likely models, black holes also respond to the
principle of simplicity, according to which among equally plausible 
models, the model involving the least number of hypotheses must be preferred. 
However, for such a wide acceptance to be settled down, the basic picture of a black hole
had to be drastically changed. The conjunction of theoretical and 
observational investigations
allowed for such a metamorphosis of the black hole image, passing from the primeval image of 
a naked black hole perfectly
passive and invisible, to the more sophisticated image of a thermodynamical engine well-feeded 
in gas and stars, which turns out to be the key of the most
luminous phenomena in the universe.
Then the modern astronomer won over to such a duality
between light and darkness may adopt the verse of the french poet L\'eon
Dierx: ``Il est des gouffres noirs dont les bords sont charmants".
 
%
%

\end{document}